\documentstyle[proceedings,namedreferences]{crckapb}

\begin{opening}
\title{A metal enriched dark cluster of galaxies at $z=1$} 
\subtitle{Dark lens searches}

\author{M.HATTORI}
\institute{Astronomical Institute, Touhoku University\\
           Aoba Aramaki, Aoba, Sendai 980-77, Japan}

\end{opening}

\runningtitle{DARK LENS SEARCHES}

\begin{document}


\section{Introduction}
Looking for and studying very distant galaxy clusters, clusters at $z>1$,
are one of the prime subjects of the modern observational 
cosmology.  
If the metallicity of the hot intra-cluster medium 
in very distant galaxy clusters is measured for example, 
it provides  fruitful  informations 
for us to understand 
the formation and evolution of galaxies.
However, difficulty of the study is that there is few confirmed very distant
galaxy clusters yet.
We first have to search for very distant clusters but
it requires very deep observations. 
A random selection of sky is not practical. 
We have to select the sky.  
In this article, it is demonstrated  that 
missing lens problem  has close connection with   
very distant cluster of galaxies and dark lens searches could  
open a new window for studying very distant cluster of galaxies. 

\section{Missing lens problem}
The word 'missing lens problem' can be found in 
the unique text book on the gravitational lensing 
written by Schneider, Ehlers, Falco (1992). 
It is the problem that there are multiple QSOs which are suspected 
to be gravitational lens systems, but for which no deflector 
is observed near the images or only a part of deflector is 
observed. The difference of the latter cases from the 
usual missing mass problem is that it requires unusually 
high mass-to-light ratio compared with that of already known objects.  
An yet unidentified part of the lens objects as already 
known objects, such as galaxy and$/$or galaxy clusters, 
is called 'dark lens'.
The dark lens search is the trial to identify 
the dark lens objects as already known objects (e.g. Wiklind \& Combes 1996). 
If all the trials in multi wave length with enough 
deepness failed to find any evidence of already known objects
as the lens, it could lead us to conclude the existence of
an yet unknown new type of object which contains few luminous matter.
 
A Table 2 summarizes the candidates of 
the multiply and ring imaged quasar and radio lenses. 
We can see that  a rather large fraction of the lens candidates 
are showing the missing lens problem.

\section{X-ray search for dark lens objects}

A dark lens object could  be a bright X-ray source since 
it may accumulate the inter-galactic medium by its strong gravitational 
field and heat the accumulated gas up to a temperature 
of a few keV 
due to the gravitational energy release,
especially for the cases of the image separation of
a few arcsec or more. 
Therefore,  
a deep pointed X-ray observations could be a promising 
method for identifying the lens objects complemental to optical
and infrared searches. 
Even if the deep X-ray observations failed to find the lens 
objects, it provides strict upper limit on the 
fraction of the hot X-ray emitting gas mass in the dark lens objects
which is one of the dominant components of luminous matter 
in the nearby  known objects. 
Non detection in X-ray, therefore, let us cast toward an 
exciting possibility, that is existence of a dark matter condensation 
containing few luminous matter. 

Motivated by those, 
we have been performing X-ray search for dark lens objects
by using the X-ray satellite ASCA and ROSAT.
In the following sections, our results obtained 
for two objects are summarized. 

\section{MG2016+112:A dark cluster} 

MG2016+112 was first discovered  as triple radio sources by Lawrence et al. (1984). 
A  part of lenses have been identified as two galaxies, galaxy D at 
$z=1.01$ (Schneider et al. 
1985 \& 1986) and galaxy C (Lawrence et al. 1984).  
However, only the luminous matter of these galaxies is not enough to explain 
the observed nature of the lensed images.  
Narasimha, Subramanian \& Chitre (1987) proposed a gravitational lensing model
that the quasar MG2016+112 is being lensed by the two galaxies  
assisted by a cluster of galaxies centered on the galaxy D.   
However, it was reported that 
no evidence of  the existence of the postulated cluster have been found
in spite of deep optical (Schneider et al. 1985) and 
infrared (Langston et al. 1991) search. 
We have performed deep X-ray observations of MG2016+112 and found evidences 
of a cluster of galaxies at $z=1$ (Hattori et al. 1997). 

\subsection{The ASCA observations}

A deep observation of MG2016+112 with the ASCA X-ray
observatory was performed from  October 24$^{\rm th}$ 
to 26$^{\rm th}$ 1994, with an effective
exposure time of 85 ksec. 
A new X-ray source (referred to
as AXJ2019+1127) was discovered in the direction of the lens system
with more than $10\sigma$ significance level.
Figure 1 shows the X-ray spectrum of AXJ2019+1127 obtained by ASCA. 
An emission line feature is clearly seen at 3.5 keV which 
matches with the redshifted Fe K line from a cluster of galaxies at $z=1$.
Assuming that the source is a cluster, the spectrum can be fit with
a  Raymond-Smith plasma model
(Raymond \& Smith 1977).  The best fitting results are summarized in Table 1. 
The inclusion of a non-zero abundance reduces $\chi^2$ by
10.01 and is significant at the 99.9\% level ($3\sigma$).
The best fit value of $N_{\rm H}$ is consistent with the Galactic value of 
$N_{\rm H}=0.15\times 10^{22}{\rm cm^{-2}}$ (Dickey \& Lockman 1990).
As shown in Fig.3, the observed X-ray luminosity and temperature 
of AXJ2019+1127 is consistent with the correlation 
of X-ray gas temperature and X-ray luminosity 
for distant clusters  
(Mushotzky \& Scharf 1997).  
These results are consistent with the idea that 
there is a cluster of galaxies including the galaxy D as 
a member galaxy.

\begin{table}[htb]
\begin{center}
\caption{The fitting results for the ASCA spectrum of AXJ2019+1127 by Raymond \& Smith model. 
Errors are the $1\sigma$ confidence level statistical errors for one interesting parameter 
except the errors for the redshift where 
the $\pm 1\%$ systematic error in the ASCA energy 
scale calibration is included in the $90\%$ statistical errors.  
The solar iron abundance of $4.67\times 10^{-5}$ is adopted 
(Anders \& Grevesse 1989) and throughout the paper except in Fig.3
$H_0=50h_{50}{\rm km/s/Mpc}$ and $q_0=0.5$ are assumed.}
\begin{tabular}{lccccc}
\hline
$kT$ & $Z$ & $L_x(2-10{\rm keV})$  &$z$ & $N_{\rm H}$ & $\chi^2/d.o.f.$\\
keV & Z$_{\odot}$ & $10^{44}h^{-2}_{50}{\rm erg/sec}$&  & ${\rm cm^{-2}}$ & \\
\hline
$8.6^{+4.2}_{-3.0}$ &  $1.7_{-0.74}^{+1.25}$&$8.4^{+2.4}_{-1.7}$ & $0.94^{+0.08}_{-0.09}$ & $0.18_{-0.10}^{+0.15}\times 10^{22}$& $46.66/52=0.90$\\
\hline
\end{tabular}
\end{center}
\end{table}

\begin{figure}
\vspace{5cm}  
\special{epsfile=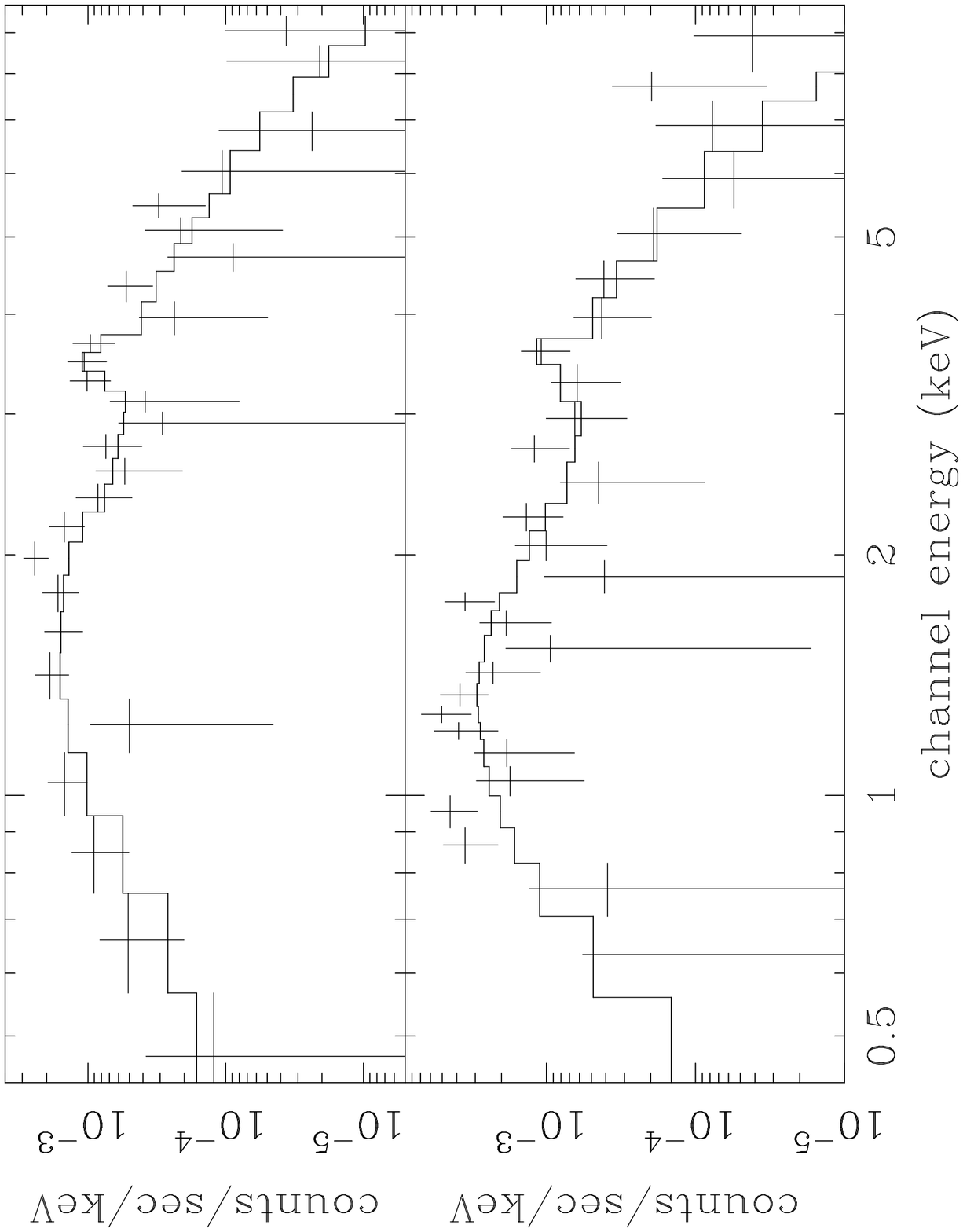 rotation=-90 hoffset=280 voffset=160 hscale=0.3 vscale=0.3}
\caption{X-ray spectrum of AXJ2019+1127 obtained by ASCA.
The background subtracted source spectrum
obtained by two GISs (top panel) and by two SISs  
(bottom panel) 
with $1\sigma$ Poisson errors
are shown.
The solid lines are the best fit Raymond-Smith models 
obtained by simultaneous fitting of the GIS 
and SIS spectra.
The background is taken from blank field
observations during the test (PV) phase of the ASCA mission.
The SIS data were taken with 2CCD bright mode observations with effective exposure time
of 22ksec. 
}
\end{figure}

\subsection{The ROSAT HRI observations}

Another crucial test for the cluster origin of the X-ray emission  
would be to   resolve the source extent. 
This was performed with a 
deep observation with the ROSAT HRI. 
We find a diffuse source with $76\pm 24$ source counts, 
($1.4 \pm 0.4) \times 10^{-3}$ cts/sec
within a circle of radius $1'$. 
Fig.2 shows the source image. 
The peak detection significance of the source is $4.2\sigma$.
The maximum of the X-ray emission is
consistent with the position of the cD galaxy D within 
the HRI  central position determination accuracy, at most
$\sim 20''$, limited both by 
pointing accuracy and poor photon statistics. 
The HRI count rate is 
consistent with, but slightly less than that expected from the
ASCA observation, that is $\sim 2.5 \times 10^{-3}{\rm cts/sec}$.  
There is no other source 
above the $3 \sigma $ significance detection limit 
in the circle of radius $3'$ centered on 
MG2016+112 in the HRI field.  
We therefore conclude  that the source detected by the 
ROSAT HRI 
is identical to AXJ2019+1127.    
Fig.2 shows that 
the X-ray surface brightness distribution of the source 
deviates significantly from the profile of the ROSAT HRI point spread function
and the source is extended. 
Since the source counts within a concentric circle monotonically 
increases up to $1'$, the source radius is estimated to be $1'$ ($500h_{50}^{-1}$kpc for $z=1$).

\begin{figure}
\vspace{5cm}  
\special{epsfile=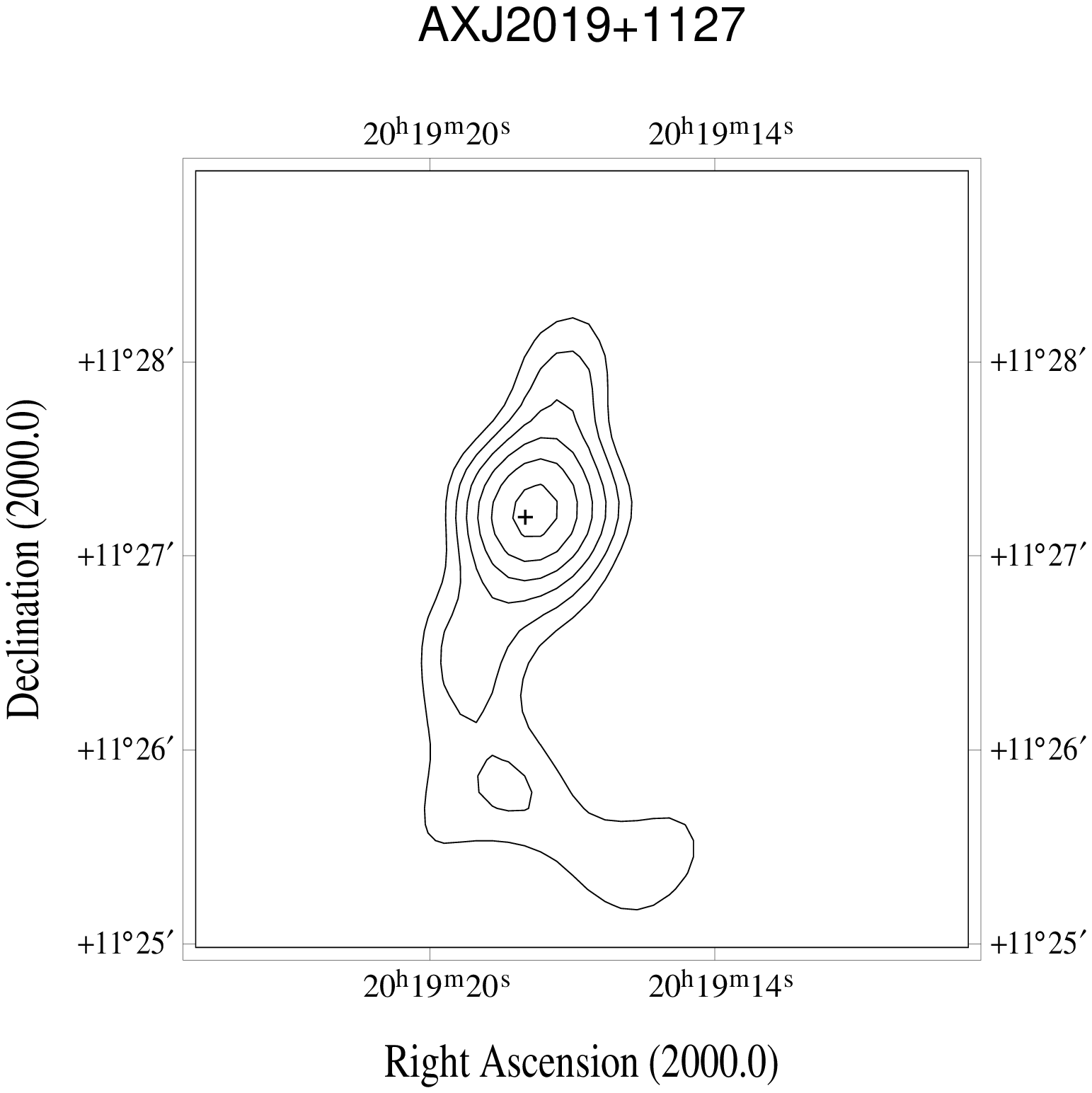 rotation=0 hoffset=0 voffset=150 hscale=0.35 vscale=0.35}
\special{epsfile=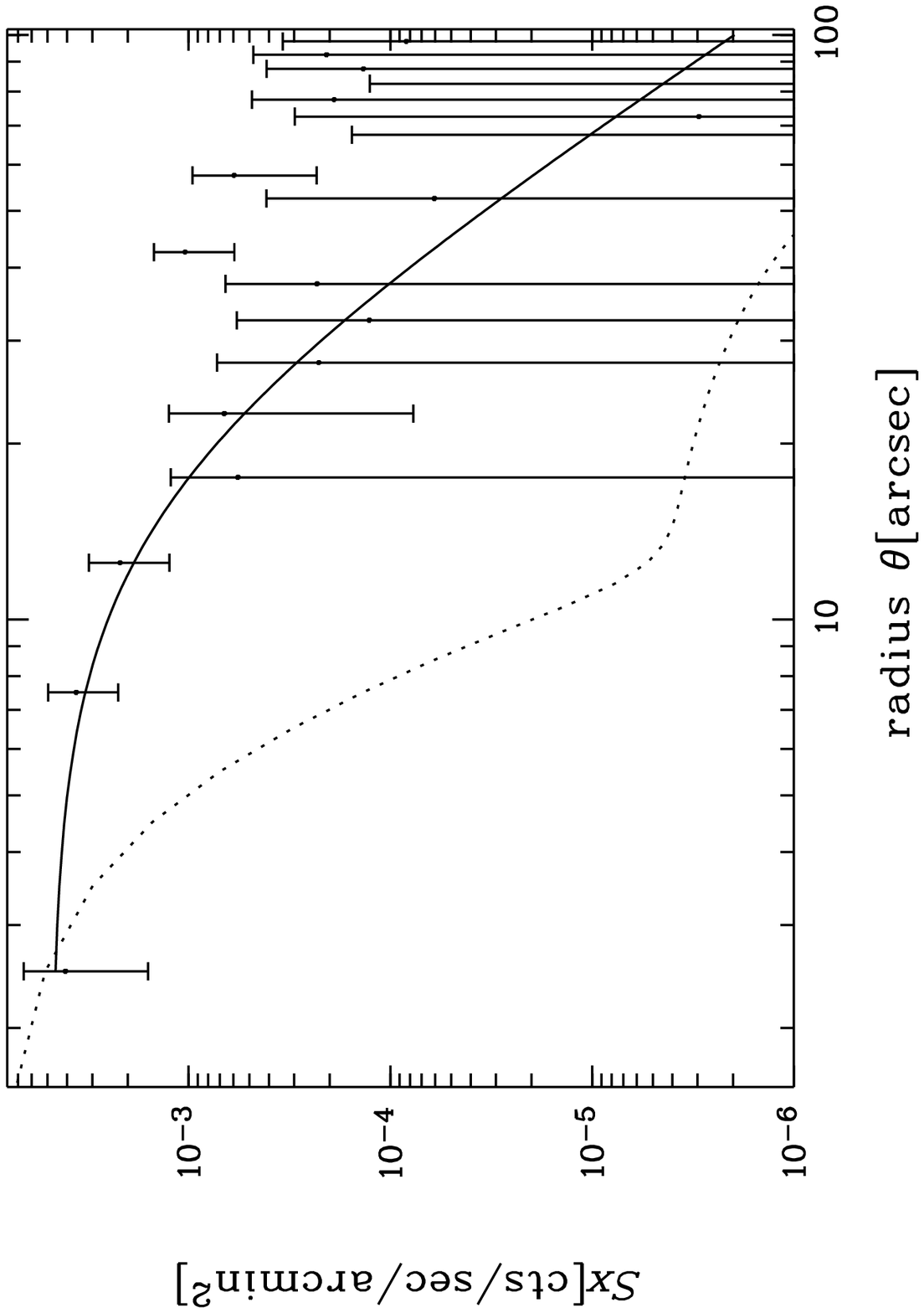 rotation=-90 hoffset=380 voffset=150 hscale=0.3 vscale=0.3}
\caption{{\bf a.} The left pannel shows that X-ray image of AXJ2019+1127 obtained by the ROSAT HRI.
A deep ROSAT HRI  observation was performed to resolve the
possible extent of the source. 
The lens system was observed from November
15$^{\rm th}$ to 20$^{\rm th}$ 1995,
from April 19$^{\rm th}$ to 28$^{\rm th}$ 1996 and
from October 22$^{\rm nd}$ to November 13$^{\rm th}$ 1996
with a total exposure time of 79 ksec, yielding
an effective exposure of 53.7 ksec after removal of observing times with
high background flux.
The accuracy of the HRI pointing is 
calibrated with two point sources coincident with stars in the Palomar
Sky Survey observed in the field. 
The accuracy of the HRI pointing is
better than $9''$.  
The contours show the significance levels with a linear
spacing of $0.5\sigma$ from the minimum level
of  $1.5\sigma$. 
The peak significance is $4.2\sigma$.
The image is smoothed
with Gaussian filter with $\sigma=15''$. 
The cross shows the
position of galaxy D.
{\bf b.} The right pannel shows that the background subtracted
 X-ray surface brightness profile of AXJ2019+1127 obtained by the ROSAT HRI.
The ROSAT HRI PSF profile at the same off-axis angle as the 
source position is
also shown (dashed line). The central surface brightness of the PSF is
set for a point source at the same central 
surface brightness as the source. 
Superposed on the data is the best fit
isothermal $\beta$ model (solid line) 
with a core radius of $17''\sim 150h_{50}^{-1}$kpc,
and $\beta=0.9$ (solid line)
which is typical for X-ray profiles 
of nearby rich clusters (Sarazin 1986). 
The central surface brightness of the model is
$\sim 3.3\times 10^{-3}{\rm cts/sec/arcmin^2}$.  
}
\end{figure}

\subsection{The most distant X-ray cluster as a dark cluster}

The above results clearly imply that AXJ2019+1127 is a galaxy
cluster in which galaxy D is the central dominant galaxy. This is the most
distant galaxy cluster discovered in X-rays so far. Since the X-ray emission
provides the best evidence that clusters are gravitationally bound 
entities (unless a large number of measured galaxy redshifts are available),
this is also the most distant clearly confirmed galaxy cluster.
The redshift of the next furthest known X-ray cluster is MS1054.5-0321 at $z=0.826$ (Donahue et al. 1997).
Assuming that the cluster is isothermal and in hydrostatic equilibrium,  
one can obtain a rough estimate of the cluster mass. The best fit 
results yields a central
electron density of $1.7\times 10^{-2}h_{50}^{0.5}$ cm$^{-3}$, a gas mass of
$2.4\times 10^{13}h_{50}^{-2.5}{\rm M_{\odot}}$ and a gravitational mass of
$3.6\times 10^{14}(kT/8.6{\rm keV})h_{50}^{-1}{\rm M_{\odot}}$  within a 
sphere of radius of
$500h_{50}^{-1}$kpc in the cluster. The gas mass fraction of 
$\sim 7h_{50}^{-1.5}\%$ of the total
mass is typical for the central regions of  nearby clusters (Briel et al. 1992)
within the measurement errors.

The obtained mass distribution is consistent with the mass model for the lens
proposed by NSC (model II in their paper). It implies that we have indeed discovered    
cluster responsible for the lensing effect additional to galaxy D.
Another spectacular result of this discovery is the high iron abundance
detected in the intra-cluster medium of AXJ2019+1127. 
Although the errors on the abundance are quite large, 
we can safely say that the iron content in this cluster 
is at least as high as that for
nearby clusters of galaxies (Ohashi 1995).  
Currently the most distant  galaxy cluster 
in which an Fe K-line has been detected is 
at a redshift of $z=0.54$ (Donahue 1996).
Therefore, AXJ2019+1127 is  now the highest redshift
cluster for which   the existence of a metal enriched intra-cluster medium 
is confirmed. 
The detection of the large iron content at high redshift sets a new 
limit for
the epoch of the enrichment.
This favors  models where early star burst phases in galaxies are
responsible for the metal enrichment of the intra-cluster 
medium (Arnaud et al. 1992; Hattori \& Terasawa 1993). 

More surprisingly, a  high iron abundance has been discovered 
in this ``dark cluster'', which is very poor in its 
galaxy content according to the deep optical and infrared searches (Schneider et al. 1985; 
Langston et al. 1991). 
The main optical light source from this cluster is galaxy D which 
has a blue luminosity of 
$L_B=1.1\times 10^{11}h_{50}^{-2}{\rm
L_{\odot}}$.  Another possible member galaxy is galaxy C (Lawrence et al. 1984) which, however,  
can contribute only a fraction of galaxy D in optical
luminosity. 
Therefore, the mass to light ratio of the cluster would be 
$M/L_B\sim  3300 (kT/8.6{\rm keV})^{-1}h_{50}M_{\odot}/L_{\odot}$!   
Since according to our current understanding the metal enrichment 
originates in the stars of the cluster galaxies, 
it is very puzzling that we detect such a high iron abundance in this
"dark cluster". 
Therefore, either AXJ2019+1127 is a very 
new and enigmatic object or it has more 
cluster members at fainter magnitudes 
that have escaped the deep searches conducted so far.
Recently, it has been reported (Ebeling et al. 1995) that there are also some low redshift 
clusters that are optically dark but bright in X-rays.
Combined with our results,  this may indicate that there exist  many such  objects 
in our universe which have eluded optical identification.  
In any case,  AXJ2019+1127 provides us with a new means for testing cosmological 
theory and ought to be well studied  
in various wavelength bands. 

\subsection{The mass distribution of the clusters at $z\sim 1$}

The radio observation has confirmed that 
there is no additional images of the 
lensed quasar MG2016+112 more than $10''$ and less than $2'$ 
from the brightest image down to two order of magnitude
fainter than the brightest image (Schneider et al. 1985).  
This gives a strong constraint on the mass
distribution of AXJ2019+1127 so that the Einstein 
ring radius of the cluster against the sources at $z=3.27$ is less than $10''$ or   $20''$ if
the possibility that the galaxy D is off-center of the cluster
within the central determination accuracy, is taken into account.
Further observations of AXJ2019+1127 to determine the center of the cluster 
with very good accuracy may provide an unique opportunity to 
study the mass distribution of this very high redshift cluster and 
test the structure formation theory (Navarro, Frenk, White 1995).

\section{Q2345+007:A cold cluster}

The quasar Q2345+007 has been one of the most mysterious lens candidate double quasar 
since its discovery (Weedman, et al. 1982). 
In spite of deep and wide field optical searches for lens object,
main lens object has not yet been 
identified (e.g. Tyson et al. 1986).   A very faint galaxy was found 
edge of the secondary image, B image,  after the subtraction of the point spread function 
of the image (Fischer et al. 1994). 
Since  a C IV doublet absorption line ($z=1.483)$ (Foltz et al. 1984, Steidel \&
Sargent 1991) was found in the B image, 
the redshift of the faint galaxy is supposed to be 
1.483. However, the expected mass of the galaxy is too small to explain
the wide separation of the two images unless it has an extremely high 
mass-to-light ratio.  
A large cluster at $z=1.49$ was claimed since both of two images have metal absorption 
lines with redshift $z=1.491$.  
Bonnet et al. (1993) reported the detection of a possible lensing cluster
from weak lensing shear field with a strength of $\gamma \sim 0.15$, confirmed 
by van Waerbeke et al. (1997)  and arclet candidates. 
After this detection, a galaxy cluster candidate was found as an enhancement 
in the number density of faint galaxies at the right location on the sky 
predicted by the shear field (Fischer et al. 1994, Mellier et al. 1994).
Pello et al. (1996) made photometric redshift estimations for the galaxies 
in the cluster candidate field and found an excess of galaxies at $z\sim 0.75\pm 0.1$.  
Assuming the cluster has this redshift, the velocity dispersion required to 
produce the observed shear pattern should be $790^{+115}_{-170}{\rm km/s}$ (Pello et al. 1996)
which corresponds to the hot gas temperature  of $kT=4.1^{+1.3}_{-1.6}$keV.
However, the cluster with this velocity dispersion centered at $40''$ off 
from the brightest image can not be the main lens for Q2345+007.

The 46.3ksec exposure of the ROSAT HRI observation and
the 80ksec ASCA observations did not detect an X-ray emission from the cluster 
responsible to the observed shear.
It sets $3\sigma$ upper limit on the X-ray luminosity of 
$0.4\times 10^{44}{\rm erg/sec}$ in rest frame $2-10$keV (Hattori et al. 1997c).
In  Fig.3, this upper limit and the temperature range expected from 
the shear field measurement are plotted.
Upper limit of the X-ray luminosity is rather small compared with the value  expected from its
 temperature. 
The cluster may  contain little hot gas. However, the errors in the temperature
estimation is large because of the simple modeling of the cluster mass distribution, 
unknown source redshifts and large errors in shear strength measurement and the cluster 
redshift estimation.  
Therefore, further observations to constrain the cluster mass
from lensing and the spectroscopical measurement of the cluster redshift are
very important. 

Up to now, dark lens searches for Q2345+007 in multi-wave length have failed  
in identifying main lens objects as galaxy and/or galaxy cluster.
It supports the idea that this double quasar is physically associated and 
is not lensed single quasar (Kochanek et al. 1997). 
Kochanek et al. (1997) is arguing that the most of the wide separation angle 
double quasars without lens could be physical association. 
However, we have to remind ourself the remarkable similarity of spectra between two images 
of the double quasars  
which is of course naturally explained by lensing, 
compared with 
the non-similarity of randomly selected two quasars' spectra 
(Simcoe, Richards, 
\& Turner 1997). 
If the most of the double quasars are turned out to be physical pair, 
it requires a new model which controls physical state 
of  two quasars  separated by  
a few$\times 10 - 100$kpc (e.g. Arp \& Crane 1992).

\begin{figure}
\vspace{7cm}
\special{epsfile=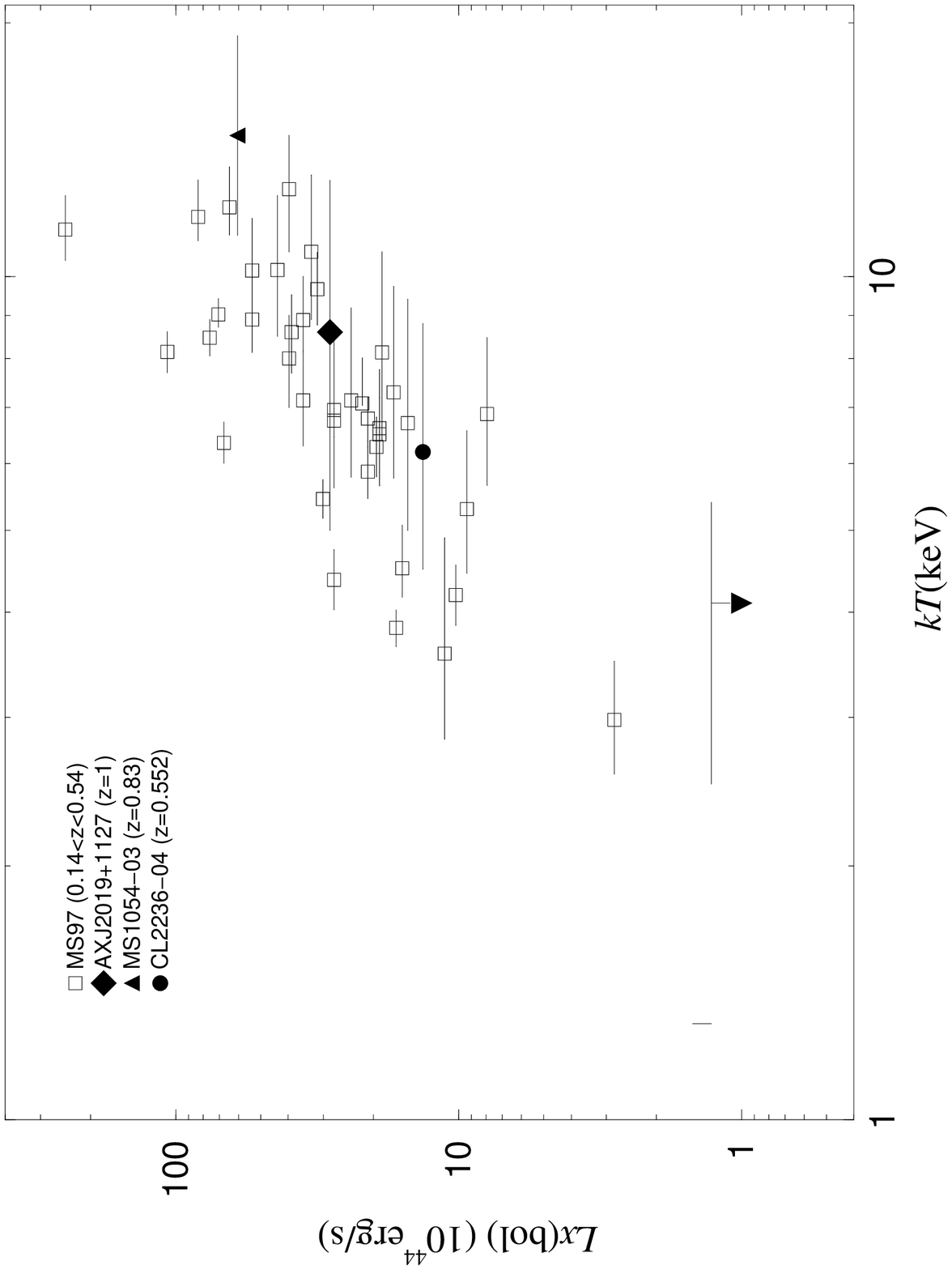 rotation=-90 hoffset=10 voffset=220 hscale=0.4 vscale=0.4}
\caption{X-ray luminosity vs gas temperature for AXJ2019+1127 and Q2345+007 cluster.
In this figure, X-ray luminosity is the bolometric luminosity 
where $q_0=0$ is assumed.  
The errors in the gas temperature are symmetrized 90\% confidence errors. The 
open squares are clusters at $0.14<z<0.54$ from Mushotzky \& Scharf (1997),
filled diamond is AXJ2019+1127, filled triangle is MS1054-03 at $z=0.83$ 
from Donahue et al. (1997), filled circle is CL2236-04 at $z=0.552$ from
Hattori et al. (1997b), and downward pointing arrow is $3\sigma$ upper limit 
of the bolometric luminosity for Q2345+007 cluster where $kT=4.5$keV is assumed
to calculate the bolometric luminosity. }
\end{figure}

\begin{table}[htb]
\begin{center}
\caption{Summary of Multiply Imaged Quasars and radio sources. From left to right, 
the name of source, the source redshift, 
the separation angle between the two main images,
the magnitude of the brightest image, 
the type of lens where blank means yet no-detection and
question mark denotes the cases 
either the detection of the lens is not secure  or unusually high mass-to-light
ratio is required, the lens redshift, the number of compact images where an E 
indicates the images are extended and an R indicates a ring, 
the degree of security as the gravitational lens where an acc denotes 
the acceptable case otherwise denoted by question mark, the reference 
for the first discovery of the source. The listed properties are mainly
quoted from Keeton \& Kochaneck (1996), Kochaneck, Falco, Mu\~noz (1997), 
and http${\rm ://www.stsci.edu/ftp/stsci/library/grav\_lens/grav\_lens.html}$.}
\begin{tabular}{lcccccccc}
\hline
Name & $z_s$ & $\Delta \theta$& $m_s$&Lens& $z_l$ &$N_{im}$&GL?& Ref\\
     &       & $''$            & mag     & &   &        &     &    \\
\hline
LBQS$2153-2056$&1.85&7.8&  &  & &2&?&$\citeauthor{hewe97}$\\
Q$2345+007$&2.15&7.3&$R=18.5$&G+CL? &1.5? 0.7? &2&?&$\citeauthor{Weed82}$\\
Q$1120+0195$&1.46&6.5&$R=15.7$&G?&$\sim 0.6?$&2&?&$\citeauthor{md89}$\\
QJ$0240-343$&1.41&6.1&$B=18.6$&  &  &2&?&$\citeauthor{tin95}$\\
Q$0957+561$&1.41&6.1&$R=16.7$&G+CL& 0.36 &2&acc&$\citeauthor{Walsh79}$\\
LBQS$1429-008$&2.08&5.1&$R=17.7$&G? $R>24$&1.5?&2&?&$\citeauthor{hewe89}$\\
MG$0023+171$&0.95&4.8&$r=22.8$&  & &2&?&$\citeauthor{hewi87}$\\
Q$2138-431$&1.64&4.5&$B_J=19.8$&$R>23.8$ &  &2&?&$\citeauthor{haw97}$\\
Q$1635+267$&1.96&3.8&$B=19.2$&G?&0.6?&2&?&$\citeauthor{ds84}$\\
MG$2016+112$&3.27&3.6&$i=22.1$&G+CL &1.01 &$>3?$ &acc &$\citeauthor{law84}$\\
3C194&1.18&3.5&$R=21.5$&G&0.31&2&?&$\citeauthor{lef88}$\\
RXJ$0911.4+0551$&2.80&3.1&$R=18.0$& & &3&acc&$\citeauthor{bade97}$\\
HE$1104-1805$&2.32&3.1&$B=16.9$&G?&1.66?&2&acc&$\citeauthor{Wis93}$\\
3C297&1.4&2.4&$R=21.0$&G& &2&?&$\citeauthor{ham90}$\\
Q$0142-100$&2.72&2.2&$R=16.8$&G&0.49&2&acc&$\citeauthor{surd87}$\\
PG1115+080&1.72&2.2&$R=15.8$&G+CL&0.29&4&acc&$\citeauthor{Wey80}$\\
MG0414+0534&2.64&2.1&$I'=19.3$&G?&1.0?&4E&acc&$\citeauthor{hewi92}$\\
CLASS1608+656&1.39&2.1&$R\sim 20$&G&0.63&4&acc&$\citeauthor{MY95}$\\
MG1131+045&1.13?&2.1& &G?&0.85?&2R&acc&$\citeauthor{ham91}$\\
MG1654+1346&1.74&2.1&$r=20.9$&G&0.25&R&acc&$\citeauthor{lang89}$\\
B1938+666&$<1.0$&1.8&$r=23$&G& &4R&acc&$\citeauthor{Pat94}$\\
Q2237+0305&1.69&1.7&$B=16.8$&G&0.039&4&acc&$\citeauthor{huc85}$\\
MG1549+3047&$>0.3?$&1.7&$R=23.3$&G&0.111&R&acc&$\citeauthor{Leh93}$\\
HE2149-2745&2.033&1.7&$B=17.3$&G?&$\sim .2-.5$&2&acc&$\citeauthor{Wis96}$\\
SBS1520+530&1.85&1.6(2.7?)&$V=18.2$&  &  &2+1?&?&$\citeauthor{Chav97}$\\
LBQS1009-0252&2.74&1.5&$B=18.1$&G?&1.62?&2&?&$\citeauthor{hewe94}$\\
CLASS1600+434&1.61&1.4&$R\sim 20$& & &2&acc&$\citeauthor{Jac95}$\\
HST14176+5226&3.41&1.4&$V=24.3$&G&0.81?&4&acc&$\citeauthor{Rat95}$\\
B1422+231&3.62&1.3&$r=15.6$&G&0.64?&4E&acc&$\citeauthor{Pat92}$\\
H1413+117&2.55&1.2&$R=17.0$&G+CL?&1.4? 1.7?&4&acc&$\citeauthor{mag88}$\\
HST12531-2914& &1.2&$V=25.5$&G&  &4&acc&$\citeauthor{Rat95}$\\
PKS1830-211& &1.0& &G&0.89 &2ER&acc&$\citeauthor{Jaun91}$\\
BRI0952-0115&4.5&0.95& & & &2&?&$\citeauthor{Mcm92}$\\
MG0751+2716&  &0.9& &G&0.35?&4R&acc&$\citeauthor{Leh94}$\\
J03.13&2.55&0.84&$R=17.1$&  &  &2&acc&$\citeauthor{clae96}$\\
Q1208+1011&3.80&0.48&$V=18.1$&  &  &2&?&$\citeauthor{mag92}$\\
B0218+357&0.96?&0.34&  &G&0.68&2ER&acc&$\citeauthor{Pat92.2}$\\

\hline
\end{tabular}
\end{center}
\end{table}

\end{document}